\documentclass[aps,prl,twocolumn,superscriptaddress,preprintnumbers,floatfix,nofootinbib]{revtex4}
\usepackage{graphicx}
\usepackage{epsfig}
\usepackage{color}
\definecolor{blue}{rgb}{0.00, 0.00, 1.00}

\definecolor{red}{rgb}{1.00, 0.00, 0.00}

\definecolor{green}{rgb}{0.0, .50, 0.00}

\begin{document}

\newcommand{\be}{\begin{eqnarray}}
\newcommand{\ee}{\end{eqnarray}}
\newcommand\del{\partial}
\newcommand{\nn}{\nonumber } 
\newcommand{\tK}{Z_{n=1}}
\newcommand{\re}{\mathrm{Re}}
\newcommand{\im}{\mathrm{Im}}
\newcommand{\Str}{\rm Str}
\newcommand{\bmat}{\left ( \begin{array}{cc} 
}
\newcommand{\emat}{\end{array} \right )}
\newcommand{\hm}{\hat{m}}
\newcommand{\hz}{\hat{z}}
\newcommand{\hx}{\hat{x}}
\newcommand{\ha}{\hat{a}}

\title{Microscopic Spectrum of the Wilson Dirac Operator} 

\author{P.H. Damgaard}
\affiliation{Niels Bohr International Academy, Niels Bohr Institute,
  Blegdamsvej 17, DK-2100, Copenhagen {\O}, Denmark} 
\author{K. Splittorff}
\affiliation{Niels Bohr Institute, Blegdamsvej 17, DK-2100, Copenhagen
  {\O}, Denmark} 
\author{J.J.M. Verbaarschot}
\affiliation{Department of Physics and Astronomy, SUNY, Stony Brook,
 New York 11794, USA}

\date   {\today}

\begin  {abstract}
We calculate the leading contribution to the spectral density of the
Wilson Dirac operator using chiral perturbation theory
where volume and lattice spacing corrections are given by universal
scaling functions. We find analytical expressions for
the spectral density on the scale of the average level spacing, 
and introduce a chiral Random
Matrix Theory that reproduces these results. Our work opens up a
novel approach to the infinite volume limit of lattice
gauge theory at finite lattice spacing and new ways to extract
coefficients of Wilson chiral perturbation theory. 
\end {abstract}

\maketitle

\noindent {\it Introduction.} 
Spectral gaps and their suppression by disorder are essential for a variety
of physical phenomena. States that intrude into the band gap, so
called Lifshitz tail states, affect the  conductivity of semiconductors
\cite{lifshitz}, they may lead to gapless superconductivity in
 superconductors with  magnetic impurities  \cite{Simons}, and they  
may show universal fluctuations given by Random
Matrix Theory (see \cite{beenrev}). Here we analyze the spectrum of 
the Wilson Dirac operator of lattice Quantum Chromodynamics (QCD). 
In the continuum limit, the Hermitian
Wilson Dirac operator has a gap equal to twice the quark mass.
At finite lattice spacing, eigenvalues of tail states intrude into the gap. 
When eigenvalues approach the center of the gap, it
becomes increasingly difficult to invert
the Wilson Dirac operator. As a consequence, such tail states can 
potentially obstruct lattice simulations. It is
therefore of
importance to have an
analytical understanding of the properties of these states.

Our results rely on two approaches, chiral
Random Matrix Theory and chiral Lagrangians
for the pseudo--Nambu-Goldstone sector of QCD.
The relation between chiral Random Matrix
Theory and the Dirac operator in theories with spontaneously
broken chiral symmetries \cite{SV} has led to a new 
understanding of the chiral limit of strongly coupled gauge theories.
The Random Matrix Theory results are universal \cite{ADMN} and are
equivalent  \cite{DOTV} to what is obtained from a 
chiral Lagrangian in the microscopic domain or $\epsilon$-regime
\cite{GL}. This gives a finite-volume scaling theory for spectral
correlation functions as well as individual eigenvalue distributions of the
continuum Dirac operator at fixed topological charge $\nu$.
In lattice QCD it has become standard to utilize these results
to obtain physical observables from simulations at finite four-volume $V$. 
There has  
for long been a desire to obtain analogous results for 
Wilson fermions at finite  lattice spacing
$a$. Here we present a solution to this problem.
  
We denote the Wilson Dirac operator by $D = D_W + m$. Below we make use
only of its block structure given by
\be
D_W = \bmat a A  & W \\ -W^\dagger & aB \emat
\label{dwil}
\ee
with $A^\dagger = A$ and $B^\dagger = B$, whereas $W$ does not have 
additional symmetry properties. The Wilson Dirac operator is 
anti-Hermitian in the continuum limit $a \to 0$ and the corresponding
eigenvalues of $D_W$ are complex away from the continuum limit. 

Although non-Hermitian, the Wilson Dirac operator satisfies 
$\gamma_5$-Hermiticity
\be
D^\dagger = \gamma_5 D \gamma_5 ~.
\label{g5Herm}
\ee
Instead of the Wilson Dirac operator itself, it is therefore often more
convenient to work with the Hermitian Dirac
operator $D_5 = \gamma_5D$. At zero lattice spacing,
the spectrum of $D_5$ has a gap around the origin of width 
 $2m$. At non-zero lattice spacing $a$, states intrude inside the
gap and for 
sufficiently large lattice spacing the gap
closes. Then one enters what is known as the Aoki phase \cite{Aoki}.
It is reminiscent of the Gorkov Hamiltonian for superconductors, where
magnetic impurities (see \cite{beenrev,Simons})
play the role of the diagonal blocks in Eq. (\ref{dwil}) and 
the Aoki phase corresponds to gapless superconductivity.
A first order scenario where the condensate jumps as a function of $m$
has been suggested \cite{SS} and support for this has been found on 
the lattice \cite{TM}.  

The spectral density $\rho_5(x)$ of $D_5$, evaluated at 
$x=0$, is an order parameter 
\cite{Heller} for the onset of the Aoki phase. 
Discretization effects in the spectrum of the Wilson Dirac operator were
analyzed by means of chiral perturbation theory in \cite{Sharpe}.
What is new here is that we obtain an exact analytical description 
in the microscopic scaling limit and show in detail the transition to
the Aoki phase. 
This opens a novel analytical approach to the
infinite-volume limit of lattice gauge theory at finite lattice spacing
and offers new ways to measure the leading coefficients of 
Wilson chiral perturbation theory. Understanding the distributions
of the low-lying
eigenvalues of the Wilson Dirac operator is also crucial for
establishing a stable domain for numerical
simulations \cite{Luscher}.

\noindent{\it Chiral Lagrangian.}
The leading-order terms of the chiral Lagrangian for Wilson fermions have been 
listed in \cite{SS}. It is a double
expansion: the continuum ordering for chiral perturbation theory 
and an expansion in the lattice spacing $a$. 
The corresponding chiral Lagrangian coincides with the continuum
Lagrangian with shifted mass plus terms starting at order $a^2$. It is
convenient  
to introduce a source for $\bar{\psi}\gamma_5\psi$, which we denote
by $z$. Here, we shall focus
on the microscopic domain where $mV$, $zV$ and
$a^2 V$ are kept fixed in the infinite-volume limit. Different counting
rules are possible \cite{Shindler}, but the present one
is most useful for elucidating the effects of finite lattice spacing
on the low-lying Dirac eigenvalues. The leading contribution to
the finite-volume QCD partition function then
reduces to a unitary matrix integral, which, up to a few
constants, is determined by symmetry arguments. 
We decompose this partition function as $Z_{N_f}=\sum_\nu Z_{N_f}^\nu$ with  
\be
Z_{N_f}^\nu(m,z;a) =  \int_{U(N_f)} \hspace{-1mm} d U \ {\det}^\nu U
~e^{S[U]} \label{Zfull}
\ee
where the action $S[U]$ for degenerate quark masses is
\be\label{lfull}
S & = & \frac{m}{2}\Sigma V{\rm Tr}(U+U^\dagger)+
\frac{z}{2}\Sigma V{\rm Tr}(U-U^\dagger)\\
&&-a^2VW_6[{\rm Tr}\left(U+U^\dagger\right)]^2
     -a^2VW_7[{\rm Tr}\left(U-U^\dagger\right)]^2 \nn\\
&&-a^2 V W_8{\rm Tr}(U^2+{U^\dagger}^2) .\nn
\ee
Below we will demonstrate that in the microscopic domain $Z_{N_f}^\nu$  
corresponds to ensembles of gauge field configurations with $\nu$ real
modes of $D_W$. The $a^2$-terms are determined by
  invariance arguments \cite{SS}, and $\Sigma$, $W_6$, $W_7$ and 
  $W_8$ are the low-energy constants of  ${\cal O}(a^2)$  
Wilson ChPT.  The two terms corresponding to
$W_6$ and $W_7$ are expected to be suppressed in the large-$N_c$ 
limit \cite{KL}, and we shall for simplicity ignore them here. 
The potential impact of these terms \cite{SharpeTopo} 
can be studied at the expense of a slightly more cumbersome
analysis. 
The leading finite-volume partition function $Z_{N_f}^\nu $ then only
depends on the microscopic scaling variables 
$\hat{m}=m\Sigma V$, $\hat{z}=z\Sigma V$, and $\hat{a}=a\sqrt{W_8V}$
which will be kept fixed for $V\to \infty$. The sign of $W_8$ will be
  discussed below.

\noindent{\it The Generating Function.}
A generating function for spectral correlation functions is
given by an average of ratios of determinants. Because of 
the inverse determinants, it has an extended graded flavor
symmetry. The graded generating function for 
spectral correlations of $D_5$ is
\be
\label{ZSUSY}
Z^\nu_{k|l}({\cal M},{\cal Z};\ha)  & = & \int \hspace{-1.5mm} dU \
{\rm Sdet}(U)^\nu \\
&& \hspace{-2.5cm}\times 
  e^{i\frac{1}{2}{\Str}({\cal M}[U-U^{-1}])
    +i\frac{1}{2}{\Str}({\cal Z}[U+U^{-1}])
    +\ha^2{{\Str}(U^2+U^{-2})}} \nn 
\ee
where ${\cal M}\equiv{\rm diag}(\hm_1\ldots \hm_{k+l})$ and 
${\cal Z}\equiv{\rm diag}(\hz_1\ldots \hz_{k+l})$.
This graded partition function differs in a subtle way from the one 
introduced in \cite{Sharpe}. As discussed in \cite{DOTV}, the 
integration manifold for non-perturbative computations is
non-compact for the bosonic sector. 
While the action (\ref{ZSUSY}) and the action
introduced in \cite{Sharpe} break the flavor symmetries in
exactly the same way, only the former one is consistent with the
convergence requirements of the graded integral for $W_8>0$. 
For perturbative calculations the convergence requirements are immaterial.  
Here and below we focus mainly on the quenched case, corresponding
to integration manifold $Gl(1|1)/U(1)$. The generalization 
to an arbitrary number of flavors is straightforward,
and we expect that the underlying integrability structure will lead to 
a full analytical solution just as in the $a=0$ case \cite{SVtoda}.

\begin{figure}[t!]
  \unitlength1.0cm
    \epsfig{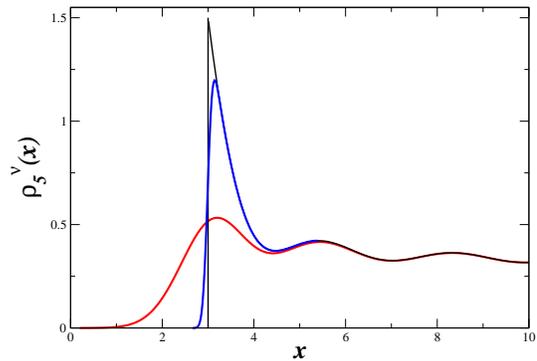}
  \caption{ \label{fig:rho5nu0} The microscopic spectrum of $D_5$
    for $\hm=3$, $\nu=0$ and $\ha=0$, $0.03$, and $0.250$. The
  $\nu=0$ spectrum is reflection symmetric about $\hx=0$.} 
  \vspace{-0mm}
\end{figure}

\begin{figure}[t!]
  \unitlength1.0cm
    \epsfig{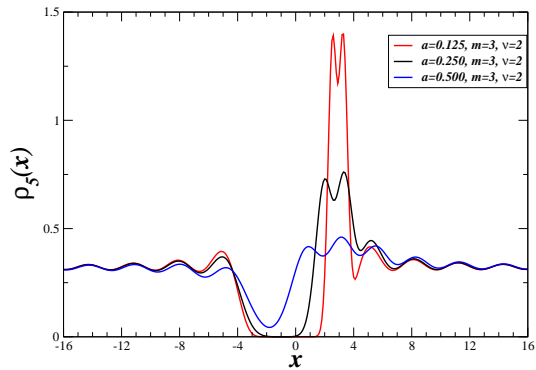}
  \caption{ \label{fig:rho5nu2} The microscopic spectrum of $D_5$
    for $\hm=3$, $\nu=2$ and $\ha=0.125$, $\ha=0.250$ and $\ha=0.500$
  respectively.}
  \vspace{-0mm}
\end{figure}

\noindent{\it The Microscopic Spectrum of $D_5$.} 
For $ a=0$, the microscopic spectral density of $D_5$ follows from 
the expression for the microscopic spectral density of $D$
through 
\be
\rho_5^\nu(\hx>\hm,\hm;\ha=0) = \frac{\hx}{\sqrt{\hx^2-\hm^2}}
\rho^\nu(\sqrt{\hx^2-\hm^2})~.
\ee 
To obtain the spectral density of
$D_5$ for $ a\ne 0$, we evaluate the resolvent 
\be
\label{G} 
G^\nu(\hz,\hm;\ha) & \equiv & 
\lim_{{\hz}'\to \hz} \frac{d}{d\hz} Z^\nu_{1|1}(\hm,\hm,\hz,{\hz}';\ha)
\ee 
and find
\be
G^\nu(\hz,\hm;\ha) & = & \int_{-\infty}^\infty ds\int_{-\pi}^\pi
\frac{d\theta}{2\pi} \ \frac i2 \cos(\theta) 
e^{S_f+S_b} 
e^{(i\theta-s)\nu} \nn \\
&& \hspace{-2cm}\times
\left(-{\hm}\sin(\theta)+i{\hm}\sinh(s)
+i{\hz}\cos(\theta)+i{\hz}\cosh(s)\right . \\
&& \hspace{-1.5cm} \left .+4{\ha}^2[\cos(2\theta)+\cosh(2s)
+(e^{i\theta+s}+e^{-i\theta-s})]+ 1\right). \nn
\ee
Here $S_f=-\hm\sin(\theta)+i\hz\cos(\theta)+2\ha^2\cos(2\theta)$ and 
$S_b=-i\hm\sinh(s)-i\hz\cosh(s)-2\ha^2\cosh(2s)$. 
As is well known, the resolvent is defined only up to ultraviolet 
subtractions in the underlying theory. The 
microscopic quenched spectral density,
\be
\rho^\nu_5(\hx,\hm;\ha) & = & \frac{1}{\pi}{\rm Im}[G^\nu(\hx,\hm;\ha))], 
\ee
is, however, uncontaminated by these ultraviolet pieces. 
Plots of $\rho^\nu_5$ are shown in Fig.~\ref{fig:rho5nu0} for $\nu=0$,
and in Fig.~\ref{fig:rho5nu2} for $\nu=2$. The eigenvalues that
converge towards the endpoints of the spectrum at ${\rm sign}(\nu) m$
in the $a \to 0$ limit are clearly visible. The sum over $\nu$ of the
spectral density, for which the continuum limit has been established
rigorously \cite{Giusti}, can be evaluated in a straightforward way.

\noindent{\it Random Matrix Theory.} An efficient alternative way to
  extend the above results to all   
spectral correlation functions and  
individual eigenvalue distributions is to construct
a chiral Random Matrix Theory that is equivalent to  the
chiral Lagrangian in the same scaling regime. This the case
if the chiral Random Matrix Theory 
has  the same global symmetries and transformation
properties as the QCD partition function with the Wilson Dirac operator.
The chiral Random Matrix Theory is 
\be
\tilde{Z}^\nu_{N_f} = \int d\tilde{A}d\tilde{B}d\tilde{W}  \  
{\det}^{N_f}(\tilde{D}_W + \tilde{m} + \tilde{z}\tilde{\gamma}_5) 
\ P(\tilde{D}_W), \:\;
\label{zwrmt}
\ee
where $\tilde{D}_W$ is of the same block form as (\ref{dwil})
and the integration is over the real and imaginary parts of the matrix
elements of the Hermitian $n \times n$ matrix, 
$\tilde{A}$, the Hermitian $(n+\nu) \times (n+\nu)$ matrix  $\tilde{B}$
and the  complex $n \times(n+\nu) $ matrix $\tilde{W}$.
This $\tilde{D}_W$ has $|\nu|$ real eigenvalues. 
We have added tildes to stress that this is a zero-dimensional matrix integral 
with parameters $\tilde{m}$ and $\tilde{z}$ instead of $m$ and $z$. In the
universal scaling limit there is a one-to-one correspondence
between the two pairs, just as in the $a=0$ case. The precise form of the
distribution of the matrix elements $P(\tilde{D}_W)$ is not important
on account of universality. The partition function (\ref{Zfull}) (with
$W_6=W_7=0$) is recovered in the microscopic scaling limit. 
For a Gaussian distribution, this can be shown by a
simple explicit calculation. It also follows that $W_8 > 0$.  
This is a consequence of the $\gamma_5$-Hermiticity:
Changing the sign of $W_8$ is equivalent to $a\to ia$, violating
$\gamma_5$-Hermiticity of $\tilde{D}_W$.
This suggests that the Hermiticity properties of the Dirac operator 
can restrict the coefficients of the effective Lagrangian. 
In fact, with $W_8<0$ the integrals in (\ref{ZSUSY}) are divergent. However,  
the graded partition function  
\be
\label{ZSUSYord}
\bar{Z}^\nu_{1|1}({\cal M},{\cal Z};\ha)  & = & \int \hspace{-1.5mm} dU \
{\rm Sdet}(U)^\nu \\
&& \hspace{-2.5cm}\times 
  e^{\frac{1}{2}{\Str}({\cal M}[U+U^{-1}])
    +\frac{1}{2}{\Str}({\cal Z}[U-U^{-1}])
    -\ha^2{{\Str}(U^2+U^{-2})}} \nn 
\ee
is now convergent.
Repeating the steps leading to (\ref{G}) with this convergent integral, 
we find a resolvent for an operator that, unlike $D_5$, is not
Hermitian. We believe that the absence of solutions in the $p$-regime
\cite{Sharpe} for $W_8<0$ has the same origin.

The ensemble with the structure of the matrix $\tilde{D}_W$ belongs to  one
of the classes in the non-Hermitian classification 
of \cite{magnea} (the $\gamma_5$-Hermiticity is there 
referred to as $Q$-symmetry). 
In the microscopic scaling limit, the partition function (\ref{zwrmt}) 
has the determinantal structure
\be
Z_{N_f}^\nu(\hm,\hz;\ha) & = &
\det[Z^{\nu+i-j}_{N_f=1}(\hm,\hz;\ha)]_{i,j=1\ldots N_f} 
\label{ZNf}
\ee
where
\be
Z_{1}^\nu & = & \int_{-\pi}^\pi \frac{d\theta}{2\pi} \ e^{i\theta\nu} 
e^{\hm\cos(\theta)+i\hz\sin(\theta)-2\ha^2\cos(2\theta)}. 
\ee
This form suggests that the partition function is a 
$\tau$-function of an integrable system of Toda type and
that an eigenvalue representation can be obtained.

The simplest quantity to compute from (\ref{ZNf}) is the chiral
condensate, $\Sigma(\hm)=\partial_{\hm}\log Z$. 
For $\nu=0$ there is a striking similarity to the chiral condensate for QCD
at non-zero isospin chemical potential $\mu$. The condensate is
constant for large $\hm$ and drops roughly linearly to zero inside a
well defined region (the Aoki phase and the pion
condensed phase, respectively).  
This is not accidental: The microscopic spectrum of $D_W$ at $a \ne 0$ forms a
thin strip along the  
imaginary axis just as the continuum Dirac operator 
does at  $\mu \ne 0$. In both cases, the
chiral condensate can be interpreted as the electric field
at $m$ of  point charges at the position of the eigenvalues.
Also the convergence requirements of the graded partition function
(\ref{ZSUSY}) have direct analogues at nonzero chemical
potential \cite{SVbos}.

\begin{figure}[t!]
  \unitlength1.0cm
    \epsfig{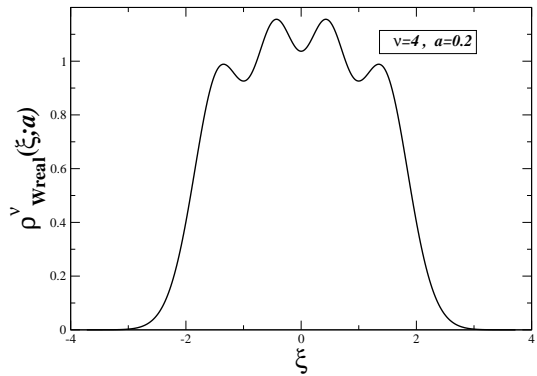}
  \caption{ \label{fig:rho-topo} The quenched density of the 
real eigenvalues of $D_W$.}
\end{figure}

\noindent
{\sl The Density of Real Modes.} 
The analytical result
for the quenched average spectral density of the real eigenvalues, denoted by 
 $\rho^\nu_{\rm Wreal}$, 
also follows from the generating function (\ref{ZSUSY}) (a derivation will
be given in \cite{DSV-II}). 
A plot of $\rho^\nu_{\rm Wreal}$ for $\nu=4$ versus $\hat
\zeta\equiv\zeta\Sigma V$ is shown in
Fig.~\ref{fig:rho-topo}. The real eigenvalues, $\zeta_i$, of $D_W$ 
repel each other and the $|\nu|=4$ real modes are clearly visible.  
We have checked that the distribution is in agreement with the chiral 
Random Matrix Theory (\ref{zwrmt}). This illustrates that in the
label $\nu$ introduced in (\ref{Zfull}) corresponds to
the number of real eigenvalues. For large $\nu$ we find that 
$\rho^\nu_{\rm Wreal}$ approaches a semi-circle. We suggest  
that analyzing just these real eigenmodes of $D_W$ may provide a new useful
tool in lattice QCD.     

\noindent{\it Distribution of Tail States.} For $|\hx-\hm|/\ha^2\gg 1$ 
and $8\ha^2 \ll 1$ the tail of the spectral density inside the gap
follows from a saddle point analysis. For $\hx>0$ we find
\be
\rho_5(\hx) \sim \exp[{-{(\hx-\hm)^2}/{16\ha^2}}],\label{rhotail}
\ee
and a similar result for $\hx < 0$.
This result applies to the tail of the blue and (marginally) of the
red curve in Fig.~\ref{fig:rho5nu0}. 
Reinstating physical parameters we find that the width 
 parameter, $\sigma$, in (\ref{rhotail}) is given by
$
\sigma^2 = {8a^2 W_8}/({V \Sigma^2})
$
so that, in the microscopic domain and sufficiently small $\ha$,
$\sigma$ scales with $1/\sqrt{V}$. Such a scaling 
has been observed for $N_f=2$ in \cite{Luscher,DelDebbio}.

Tail states may also be studied in their own right and for
applications in condensed matter physics. In particular,
in the thermodynamic limit, $\hm, \hx, \ha^2 \gg 1$,
the average level density of $D_5$ can be obtained by a saddle point
analysis. For $8\hat a^2/\hat m < 1$ it
 vanishes inside $[-\hx_c, \hx_c]$ with $\hx_c$ given by
$
\hx_c = 8 \hat a^2\left [ ( \hat m/{8\hat a^2}\right)^{2/3}
-1]^{3/2}.
$  
It has exactly the same form as for superconductors with
magnetic impurities \cite{Simons}. 
In the scaling limit where $V^{2/3} (x_c -x)$ is kept fixed,
the spectral density can be computed inside the gap by a saddle
point approximation of (\ref{ZSUSY}), and it agrees with
universal Random Matrix Theory results for the so-called soft edge.
Such universal behaviour has also been found in condensed matter
systems \cite{Simons,beenrev}. 

\noindent {\it Conclusions.} Using a graded 
chiral Lagrangian for Wilson fermions at finite lattice spacings, we 
have obtained an analytical form for 
the Wilson Dirac spectrum at fixed number, $\nu$, of real
eigenvalues. These
results, and their extensions to dynamical fermions, 
should be useful for lattice simulations at finite volume. 
We have shown how the leading low-energy constant for Wilson fermions,
$W_8$, can be extracted from lattice spectra of the
Wilson Dirac operator in the  $\epsilon$-regime. 

The problem can also
be reformulated in terms of a new chiral Random Matrix Theory that 
describes spectral correlation functions of the Wilson Dirac
operator in the appropriate scaling regime. These results 
%have interesting analogies in systems of condensed matter physics, and
%they 
open  a new domain of Random Matrix Theory where chiral
ensembles merge with Wigner-Dyson  ensembles. Essential to this
study is an analysis of the gap of the Wilson Dirac operator at 
finite mass. Lattice QCD simulations depend crucially
on control of this gap and its variation as a function of the lattice spacing. 
As we have stressed, our results are not only important for 
understanding the finite lattice spacing effects
of Wilson fermions near the chiral limit, 
but may have interesting applications to
tail states in condensed matter systems.

\noindent
{\bf Acknowledgments:}
This work was supported by U.S. DOE Grant No. DE-FG-88ER40388 (JV) and the 
Danish Natural Science Research Council (KS).  We thank B. Simons, 
M. L\"uscher and P. de Forcrand for discussions.

\end{document}